\documentclass[11pt,twocolumn]{article} 

\usepackage{graphicx}
\usepackage{mathenv}
\usepackage{latexsym} 
\usepackage{amssymb,amsmath,amsthm}
\usepackage[T1]{fontenc}
\usepackage{verbatim}
\usepackage[english]{babel}    
\usepackage{vmargin}      
\usepackage{mathenv}
\usepackage{alltt}
\usepackage{color}
\usepackage{abstract}
\newcommand{\deriv}{\mathrm{d}}
\renewcommand{\exp}{\mathrm{e}}

\title{Flute-like musical instruments: a toy model investigated through numerical continuation}

\author {S. Terrien$^{a,\ref{l1}}$, C. Vergez$^{a}$, B. Fabre$^{b}$\\
\\
$^a$ Laboratoire de Mécanique et d'Acoustique (LMA-CNRS UPR 7051),\\ Aix Marseille Université,\\
31 chemin Joseph Aiguier, 13402 Marseille cedex 20, France\\
\\
$^b$ LAM, Institut Jean Le Rond d'Alembert (CNRS UMR 7190),\\
 UPMC Univ. Paris 6,\\
11 rue de Lourmel, 75015 Paris, France }
\date{}

\begin{document}
\twocolumn[
\maketitle
\begin{onecolabstract}

Self-sustained musical instruments (bowed string, woodwind and brass instruments) can be modeled by nonlinear lumped dynamical systems.
Among these instruments, flutes and flue organ pipes present the particularity to be modeled as a delay dynamical system. In this paper, such a system, a toy model of flute-like instruments, is studied using numerical continuation. Equilibrium and periodic solutions are explored with respect to the blowing pressure, with focus on amplitude and frequency evolutions along the different solution branches, as well as "jumps" between periodic solution branches. The influence of a second model parameter (namely the inharmonicity) on the behaviour of the system is addressed. It is shown that harmonicity plays a key role in the presence of hysteresis or quasi-periodic regime. Throughout the paper, experimental results on a real instrument are presented to illustrate various phenomena, and allow some qualitative comparisons with numerical results.

\vspace{5mm}

\end{onecolabstract}
]

\vspace{5mm}
\section{Introduction}
\label{Intro}

\footnotetext[1]{\label{l1}Corresponding author: \texttt{terrien@lma.cnrs-mrs.fr}}

Sound production in self-sustained musical instruments, like bowed string instruments, reed instruments or flutes, involves the conversion of a quasi-static energy source (provided by the musician) into acoustic energy.
This generation of auto-oscillations necessarily implies nonlinear mechanisms. Thus, even the simplest models of self-sustained musical instruments should include nonlinear terms \cite{McIntyre_83}.

We focus in this paper on flute-like instruments. Many works with both experimental and modeling approaches have highlighted the wide variety of their oscillation regimes (for example \cite{Verge_97,Coltman_06}), and aim to explore the complexity of their dynamics \cite{Fletcher_76,Schumacher_78,Auvray_12}. Although different studies \cite{McIntyre_83,Coltman_92,Coltman_06} has dealt with numerical investigation of the behaviour of simple models of flute-like instruments, no study, to the authors knowledge, has ever investigated the determination of solution branches using numerical continuation, and the analysis of jumps between branches. This is done in this paper.

A numerical continuation approach has recently proved to be useful in the context of musical instruments, especially for reed instruments (like the clarinet) \cite{Karkar_11}. However, unlike reed instruments, flute-like instruments are delay dynamical systems (see section \ref{section_systeme}), which complicates the model analysis, and prevents the use of the same numerical tools (as, for example, Auto \cite{Doedel_81} or Manlab \cite{Cochelin_09}).

Using a numerical continuation software dedicated to delay dynamical systems \cite{manuel_Biftool}, we study in this paper a dynamical system inspired by the physics of flute-like instruments. We call it a toy model, since it is known in musical acoustics that more accurate (much more complex) models should be considered \cite{Chaigne_Kergomard,Fabre00}. We investigate throughout the paper the diversity of oscillation regimes of the toy model, as well as bifurcations and jumps between branches. We particularly focus on the nature of the solutions (static or oscillating), and in the later case, on the evolution of frequency and amplitude along periodic solution branches. Indeed, in a musical context, periodic solutions correspond to notes, and oscillating frequency and amplitude are respectively related to the pitch and the intensity of the note played. Comparisons between the obtained bifurcation diagrams, experimental data and time-domain simulations highlight the valuable contribution of numerical continuation. Moreover, surprisingly enough, the qualitative behaviour of the toy model displays close similarities to that of a real instrument.

The paper is structured as follows. The dynamical system under study is presented in section \ref{section_systeme}. In section \ref{section_sol_perio}, we present the results of numerical continuation of the branches of static and periodic solutions and their stability analysis, with focus on amplitude and frequency evolutions. The study of "jumps" between different periodic solution branches is presented in section \ref{section_bif_br_perio}. Finally, we discuss, in section \ref{section_reg_QP}, the influence of a parameter related to the instrument makers' work on the characteristics of the different oscillating regimes. Qualitative comparisons with the sound produced by real instruments and with the experience of an instrument maker are presented.




\section{System studied (toy model)}
\label{section_systeme}

In this paper, we study the following delay dynamical system:

\begin{System}
z(t) = v(t-\tau)\\
p(t) = \alpha \cdot \tanh\[z(t)\]\\
V(\omega) = Y(\omega) \cdot P(\omega)
\label{systeme_eq_modele}
\end{System}

\noindent
where lowercase variables are written in the time domain, and uppercase variables in the frequency domain. The unknowns are $z$, $v$, and $p$, while $\alpha$ and $\tau$ are scalar constant parameters. $Y$ is a known function of the frequency, and will be detailed later. We first briefly describe the mechanism of sound production in flute-like instruments, and then we precise to which extent this system can be interpreted as an extremely simplified model of this kind of musical instruments.

\subsection{Sound production in flute-like instruments}

\begin{figure}
\centering
\includegraphics[trim = 0cm 5cm 0cm 0cm, clip=true,width=\columnwidth]{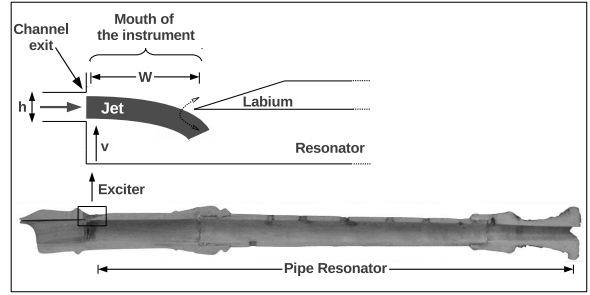}
\caption{Recorder section, and simplified representation of the jet oscillation around the labium, in the mouth of the resonator.}
\label{schema_coupe_flute}
\end{figure}

Since the work of Helmholtz \cite{Helmholtz}, a classical approach consists in modeling a musical instrument by a nonlinear coupling between an exciter and a resonator. In other wind instruments, the exciter involves the vibration of a solid element: a cane reed in the clarinet or the saxophone, the musician's lips in the trumpet or the trombone... In flute-like instruments (which includes recorders, transverse flutes, flue organ pipes...), the exciter consists in the oscillation of the blown air jet around an edge called labium (see figure \ref{schema_coupe_flute}) \cite{Rayleigh,Fabre00}.

The jet-labium interaction generates the pressure source which excites the resonator formed by the pipe, and thus creates an acoustic field in the instrument. In turn, the acoustic field disturbs the jet at the channel exit (see for example \cite{Coltman_92b}). As the jet is naturally unstable, this perturbation is convected and amplified along the jet \cite{Rayleigh, Nolle_98}, from the channel exit to the labium, which sustains the oscillations of the jet around the labium, and thus the sound production \cite{Chaigne_Kergomard, Fabre00}.
\\

The convection time of perturbations along the jet introduces a delay in the system, whose value is related to the convection velocity $c_v$ of these hydrodynamic perturbations. Theoretical \cite{Rayleigh, Nolle_98} and experimental \cite{De_La_Cuadra_07} studies have shown that  $c_v$ is related to the jet velocity $U_j$ (itself related to the blowing pressure applied by the musician) through:
\begin{equation}
c_v \approx 0.5 \cdot U_j.
\end{equation}

\noindent
Noting $W$ the mouth height (\textit{i.e.} the length of the jet, between the channel exit and the labium, highlighted in figure \ref{schema_coupe_flute}), an approximation of the delay value is given by:

\begin{equation}
\tau = \frac{W}{c_v} \approx \frac{W}{0.5 \cdot U_j}.
\label{equation_retard}
\end{equation}

\subsection{System studied: a toy model of flute-like instruments}

\begin{figure}
\centering
\includegraphics[width=\columnwidth]{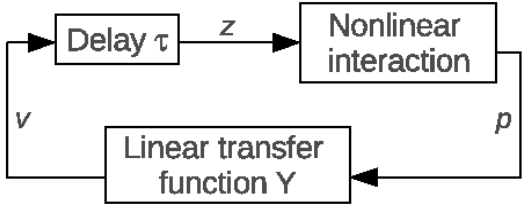}
\caption{Basic modeling of sound production mechanism in flute-like instruments, as a system with a feedback loop \cite{Chaigne_Kergomard,Fabre00}.}
\label{schema_syst_boucle}
\end{figure}

This mechanism of sound production can be modeled by a nonlinear oscillator, such as the one represented in figure \ref{schema_syst_boucle} \cite{Chaigne_Kergomard,Fabre00}. This very basic modeling includes the three following elements:
\begin{itemize}
\item{a delay, related to the convection time of the hydrodynamic perturbations along the jet.}
\item{a nonlinear function, representing the jet-labium interaction.}
\item{a linear transfer function representing the acoustic behaviour of the resonator.}
\end{itemize}

\noindent
Therefore, system (\ref{systeme_eq_modele}) can be seen as a very basic model of flute-like instruments, and thus can be related to the nonlinear oscillator represented in figure \ref{schema_syst_boucle}. Variables $z$, $p$ and $v$ are then respectively related to the transversal deflection of the jet at the labium, to the pressure source created by the jet-labium interaction, and to the acoustic velocity induced by the waves at the pipe inlet, so-called "mouth of the instrument" (see figure \ref{schema_coupe_flute}). The scalar parameters $\alpha$ and $\tau$ are respectively associated to the amplification of the perturbations along the jet, and to the convection delay defined by equation (\ref{equation_retard}).
\\

Using a modal decomposition of the resonator acoustical response, the transfer function is written in the frequency domain as a sum of $m$ modes (as it is done, for example, in the work of Silva \cite{Silva_07}):%

\begin{equation}
Y(\omega) = \sum_{k=1}^m \frac{a_k \cdot \mathrm{j}\omega}{\omega_k^2-\omega^2+\mathrm{j}\omega \cdot \frac{\omega_k}{Q_k}}
\label{eq. admittance}
\end{equation}

\noindent
where $a_k$, $\omega_k$ and $Q_k$ are respectively the modal amplitude, the resonance pulsation and the quality factor of the k-th mode.
\\

This representation is very simplified compared to the most recent models of flute-like instruments \cite{Chaigne_Kergomard,Fabre00}, which take into account the following elements:

\begin{itemize}
\item{a precise modeling of the aeroacoustic source, which includes, in the latest models, a time derivative of the delayed term \cite{Fabre00}.}
\item{nonlinear losses, due to vortex shedding at the labium \cite{Fabre_96}.}
\item{an accurate description of the resonator: in theory, the modal decomposition in equation (\ref{eq. admittance}) can include a large number $m$ of modes. For sake of simplicity, we only retain in this paper at most the first two modes of the pipe.}
\end{itemize}

Although this "toy model" may seem similar to those studied by McIntyre et al. \cite{McIntyre_83} and Coltman \cite{Coltman_92}, an important difference relies on the description of the resonator: we use here a modal decomposition, whereas the models of McIntyre et al. \cite{McIntyre_83} and Coltman \cite{Coltman_92} are both based on a time-domain description of the resonator behaviour.

Here, the interest of using a modal description is the ability to change independently of each other the different resonator parameters (such as, for example, the resonance frequencies - see section \ref{section_reg_QP}).

\subsection{Rewritting as a first-order system}

To be analysed with classical numerical continuation methods for delay differential equations \cite{Engelborghs_02}, system (\ref{systeme_eq_modele}) should be rewritten as a classical first-order delay system $ \mathbf{\dot x}(t) = f(\mathbf{x}(t),\mathbf{x}(t-\tau),\boldsymbol{\gamma})$. In such a formulation, $\mathbf{x}$ is the vector of state variables and $\boldsymbol{\gamma}$ the set of parameters. To improve numerical conditioning of the problem, a dimensionless time variable and a dimensionless convection delay are defined : 
\begin{System}
\tilde t = \omega_1 t \\
\tilde \tau = \omega_1 \tau
\label{syst_adim_t}
\end{System}
with $\omega_1$ the first resonance pulsation (see equation (\ref{eq. admittance})). Rewritting system (\ref{systeme_eq_modele}) finally leads to the following system of $2m+2$ equations, where $m$ is the number of modes in the transfer function $Y(\omega)$ (see \ref{annexe_Biftool} for more details): 
\newline
$\forall$ $i=[1, ..., m]$ (where $i$ is an integer),

\begin{System}
\begin{split}
v(\tilde t) =& \sum_{k=1}^m v_k(\tilde t)\\
y(\tilde t) =& \sum_{k=1}^m y_k(\tilde t)\\  
\dot v_{i}(\tilde{t}) =& y_i(\tilde{t})\\
\dot y_i (\tilde{t}) =& \frac{\alpha a_i}{\omega_1} \{ 1-\tanh^2\left[ v(\tilde{t}-\tilde{\tau}) \right] \} \cdot y(\tilde{t}-\tilde{\tau}) \\
& - \left( \frac{\omega_i}{\omega_1} \right)^2 v_{i}(\tilde{t})  - \frac{\omega_i}{\omega_1 Q_i} y_i(\tilde{t}).  
\end{split}
\label{eq:1st-order}
\end{System}
System (\ref{eq:1st-order}) is studied throughout this paper.
Numerical results are computed using the software DDE Biftool \cite{manuel_Biftool}, which performs numerical continuation of delay differential equations using a prediction/correction approach \cite{Engelborghs_02}. Periodic solutions are computed using orthogonal collocation \cite{Engelborghs_00}. Typically, we use about 100 points per period, and the degree of the Lagrange interpolation polynomial is 3 or 4.




\section{Periodic regimes emerging from the equilibrium solution}
\label{section_sol_perio}
\subsection{Branch of equilibrium solutions}
\label{sections_sol_statiques}
In flute-like instruments, the delay $\tau$ is related to the jet velocity (see equation (\ref{equation_retard})), and therefore to the pressure into the musician's mouth through the  Bernoulli equation for stationary flows. To analyze the system (\ref{eq:1st-order}), it is thus particularly interesting to choose this variable as the continuation parameter.

Regardless of the parameter values, it is obvious that system (\ref{eq:1st-order}) has a unique equilibrium solution defined by:
\begin{System}
\forall i=[1, 2, ..., m] :\\
v_i(\tilde t) = 0\\
y_i(\tilde t) = 0.
\label{equilibrium_sol}
\end{System}

A stability analysis of this static solution is performed both through computation of the eigenvalues of the linearized problem and the analysis of the open-loop gain (see \ref{annexe_AL} for more details).
Moreover, this linear analysis allows to distinguish two kinds of emerging periodic solutions:
\begin{itemize}
\item{those resulting from the coupling between an acoustic mode of the resonator (a particular term in the sum defining the transfer function $Y(\omega)$ in equation (\ref{eq. admittance})) and the first hydrodynamic mode of the jet (corresponding to $n=0$ in equation (\ref{equation_GBO}) of \ref{annexe_AL}).}
\item{those resulting from the coupling between an acoustic mode of the resonator and an higher-ranked hydrodynamic mode of the jet (corresponding to $n>0$ in equation (\ref{equation_GBO})).}
\end{itemize}

Indeed, for flute-like instruments, an auto-oscillation results from the coupling between an acoustic mode of the resonator and an hydrodynamic mode of the jet \cite{Verge_97,Chaigne_Kergomard}. As also observed experimentally by Verge \cite{Verge_97} in the case of a flue organ pipe, the first case $n=0$ corresponds to the "standard" regimes in flute-like instruments. On the contrary, the second case $n>0$ corresponds to the so-called "aeolian" regimes, for which the time delay $\tau$ is larger than the oscillation period $\frac{2\pi}{\omega}$ of the system (see equation \ref{equation_GBO} in \ref{annexe_AL}). In the context of musical instruments, it corresponds to particular sounds, like for example those produced when the wind machine of an organ is turned off leaving a key pressed.
\\

We focus here on these two kinds of periodic solutions, and we study particularly the amplitude and frequency evolutions along the branches.

\subsection{Branches of periodic solutions: aeolian and standard regimes}
\label{section_reg_periodiques}

For sake of clarity, we consider in this section a transfer function containing a single acoustic mode (\textit{i.e.} $m=1$ in the expression of $Y(\omega)$, defined by equation (\ref{eq. admittance})). The addition of a second acoustic mode will be discussed in the next section.

\begin{figure}
\begin{minipage}[c]{.99\columnwidth}
\centering
\includegraphics[width=\columnwidth]{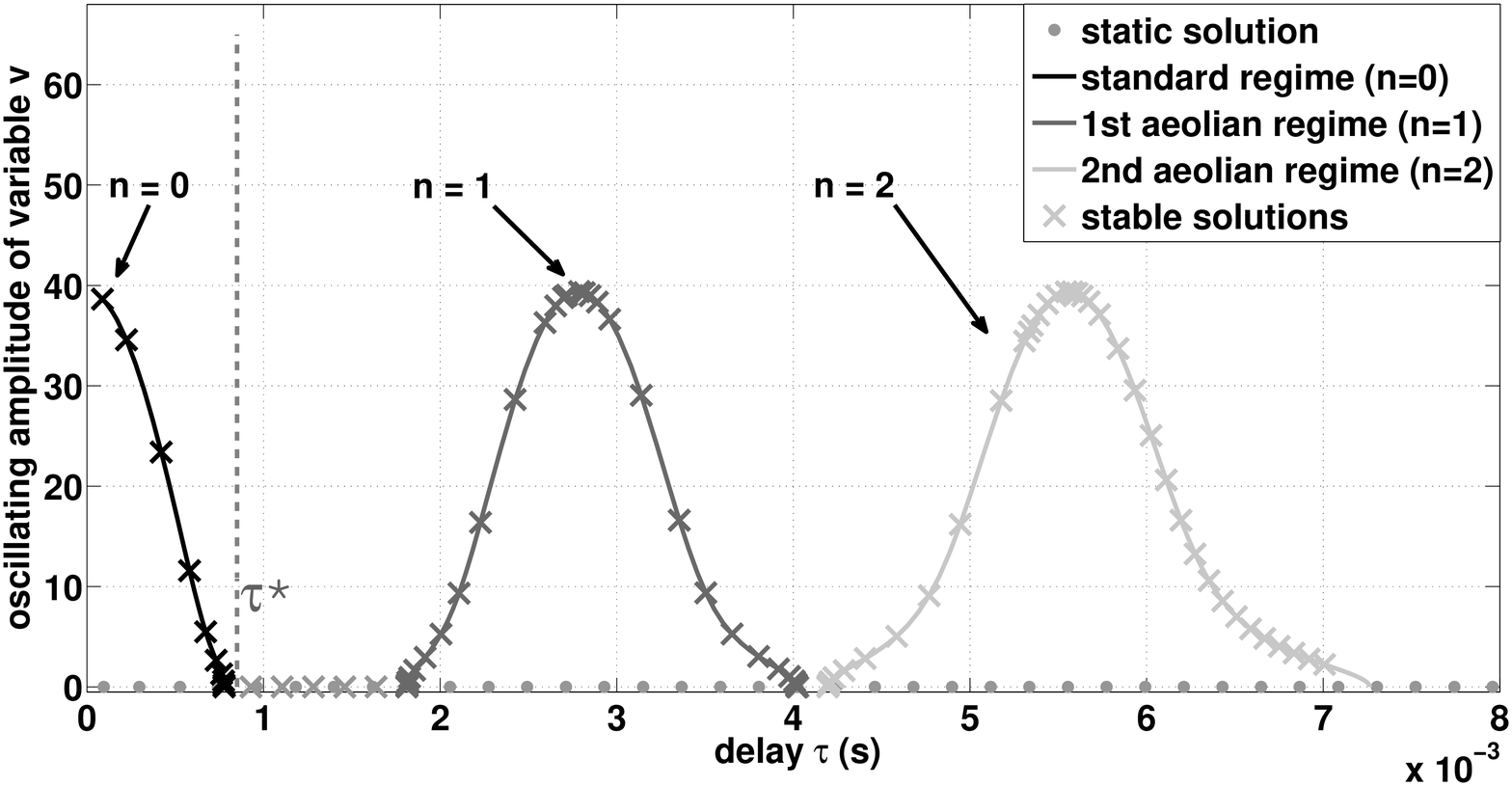}
\caption{Bifurcation diagram of system (\ref{eq:1st-order}).
Abscissa: delay $\tau$ (in seconds). Ordinate: amplitude of the oscillating variable $v(t)$. Symbols $\times$ indicate the stable parts of the branches, and the dashed line indicate the separation between the standard regime and the aeolians ones. Parameters value: $m = 1$, $\alpha = 10$ ; $a_1 = 70$ ; $\omega_1 = 2260$ ; $Q_1 = 50$. Values of $n$ are computed using equation (\ref{equation_GBO}) in \ref{annexe_AL}. } 
\label{diag_bif_tau_dim}
\end{minipage}
\end{figure}

Numerical continuation of the different periodic solution branches allows to display the bifurcation diagram showed in figure \ref{diag_bif_tau_dim}, representing the amplitude of the oscillating variable $v(\tilde t)$ defined in system (\ref{eq:1st-order}), with respect to the delay $\tau$. It is useful to keep in mind that blowing harder makes the jet velocity $U_j$ increase and thus $\tau$ decrease (according to equation (\ref{equation_retard})).

The stability of each branch is addressed by computing the Floquet multipliers \cite{Engelborghs_02}: since all of them lie within the unit circle, it can be concluded that all the branches are stable \cite{Nayfeh}.

As shown in figure \ref{diag_bif_tau_dim}, the use of equation (\ref{equation_GBO}) (\ref{annexe_AL}) allows to distinguish the different kind of periodic regimes, and highlights that the only standard regime (related to the first hydrodynamic mode of the jet) is located in the part of the graph where $\tau$ is lower than $\tau* = 0.9\cdot10^{-3}$ s. The other branches correspond to aeolian regimes related to the second and the third hydrodynamic modes of the jet (respectively $n=1$ and $n=2$). 

A larger range in $\tau$ in figure \ref{diag_bif_tau_dim} would reveal other branches of aeolian sounds. They correspond to the infinite series of aeolian instabilities highlighted for each acoustic mode of the transfer function $Y(\omega)$ in \ref{annexe_AL}.

\begin{figure}
\begin{minipage}[c]{.99\columnwidth}
\centering
\includegraphics[width=\columnwidth]{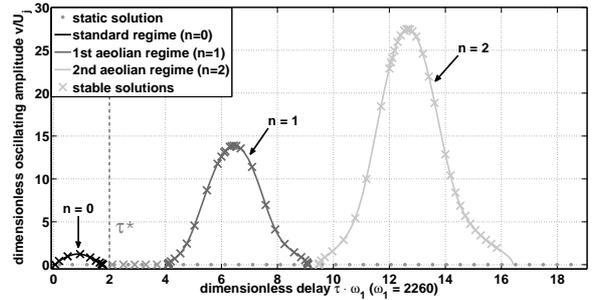}
\caption{Bifurcation diagram of system (\ref{eq:1st-order}).
Abscissa: dimensionless delay $\tilde \tau = \omega_1 \tau$. Ordinate: oscillation amplitude of the dimensionless variable $v(\tilde t)/U_j$. Symbols $\times$ indicate the stable parts of the branches, and the dashed line indicate the separation between the standard regime and the aeolians ones.. Parameters value: $m = 1$, $\alpha = 10$ ; $a_1 = 70$ ; $\omega_1 = 2260$ ; $Q_1 = 50$. Values of $n$ are computed using equation (\ref{equation_GBO}) in \ref{annexe_AL}.} 
\label{diag_bif_tau}
\end{minipage}
\end{figure}

As highlighted previously, numerical calculations are performed using the dimensionless parameter $\tilde \tau$ (see equations \ref{syst_adim_t}). Throughout the paper, for sake of consistency, we thus present the results using $\tilde \tau$ and the dimensionless amplitude $v(\tilde t)/U_j$ of the variable $v(\tilde t)$ (with $U_j$ the jet velocity). 
Thereby, figure \ref{diag_bif_tau} represents the same bifurcation diagram as figure \ref{diag_bif_tau_dim}. 
It is important to keep in mind that the jet velocity $U_j$ is not a constant (see equation (\ref{equation_retard}) for the relation between $U_j$ and the continuation parameter $\tau$): although the order of magnitude of the dimensionless amplitude of the different regimes seems to be different (figure \ref{diag_bif_tau}), it corresponds to a same value of the dimensioned amplitude (figure \ref{diag_bif_tau_dim}).

\subsubsection{Amplitude evolutions}

To study flute-like instruments, it is useful to define the dimensionless jet velocity $\theta$ \cite{Meissner_01,Verge_97}: 
\begin{equation}
\theta = \frac{U_j}{W \cdot f}
\label{def_theta}
\end{equation}
where $W$ is the jet length (see figure \ref{schema_coupe_flute}), and $f$ the oscillation frequency.

\begin{figure}
\centering
\includegraphics[width=\columnwidth]{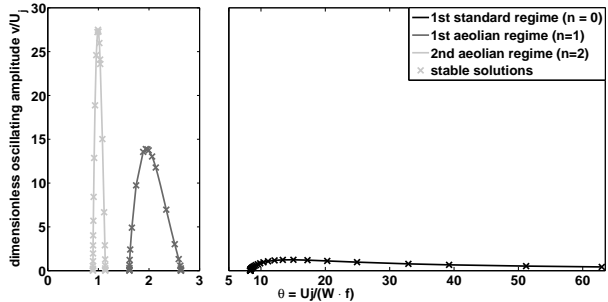}
\caption{Bifurcation diagram of the system (\ref{eq:1st-order}) with same parameter values as in figure \ref{diag_bif_tau}.
Abscissa: dimensionless jet velocity $\theta$. Ordinate: oscillation amplitude of the dimensionless variable $v(\tilde t)/U_j$. Symbols $\times$ indicate the stable parts of the branches.}
\label{figure_ampl_theta}
\end{figure}

Indeed, the representation, in figure \ref{figure_ampl_theta}, of the oscillation amplitude of the dimensionless variable $v(\tilde t)/U_j$ (related, in flute-like instruments, to the acoustic velocity in the mouth of the resonator) as a function of $\theta$ for each branch of periodic solutions plotted in figure \ref{diag_bif_tau} shows a clear separation between aeolian and standard regimes in two different zones of the dimensionless jet velocity $\theta$. This is interesting since for flute-like instruments playing under normal conditions,  $\theta$ is expected to be larger than  4 \cite{These_Blanc}. Figure \ref{figure_ampl_theta} shows the same trend for the standard regime. On the contrary, we observe that all aeolian regimes are found for $\theta < 4$. This feature is in agreement with the fact that they occur at low jet velocities. Indeed, for the same oscillation frequency, $\theta$ is lower for an aeolian regime than for a standard regime (as highlighted in section \ref{section_sol_perio}, $\tau f$ is larger than unity for aeolian regimes).

Moreover, the bifurcation diagram displayed in figure \ref{figure_ampl_theta} shows a particular amplitude evolution for aeolian regimes. Indeed, one notices, for such regimes, a bell-shaped curve, whereas for the "standard" regime one can observe a saturation followed by a slow decrease of the amplitude. 
Thus, while the standard regime exists when $\tau$ tends to zero (and thus when the jet velocity $U_j$ tends to infinity - see equation (\ref{equation_retard})), aeolian regimes conversely exist only for restricted ranges of these two parameters.

\subsubsection{Frequency evolutions}

\begin{figure}
\centering
\includegraphics[width=\columnwidth]{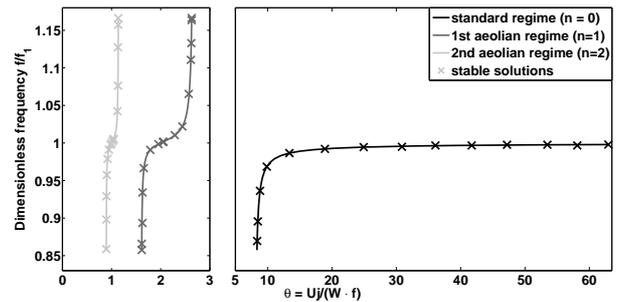}
\caption{Bifurcation diagram of the system (\ref{eq:1st-order}), with same parameter values as in figure \ref{diag_bif_tau}.
Abscissa: dimensionless jet velocity $\theta$. Ordinate: dimensionless oscillation frequency $\frac{f}{f_1}$ (with $f_1 = \frac{\omega_1}{2\pi}$ the resonance frequency of the transfer function $Y(\omega)$). Symbols $\times$ indicate the stable parts of the branches.}
\label{figure_se_freq}
\end{figure}

Figure \ref{figure_se_freq} presents the dimensionless oscillation frequency $f/f_1$ (with $f_1 = \frac{\omega_1}{2\pi}$ the resonance frequency of the transfer function $Y(\omega)$) as a function of $\theta$, along the different periodic solution branches shown in figure \ref{diag_bif_tau}. Similarly to the amplitude, it reveals different evolution patterns for aeolian and standard regimes:
\begin{itemize}
\item For the standard regime, one observes an important frequency deviation just after the oscillation threshold (at $\theta = 8.3$). Then the frequency asymptotically tends toward the resonance frequency  $f_1$ as $\theta$ increases further.
\item For aeolian regimes, one observes an important frequency deviation  just after the oscillation threshold, followed
by a second frequency deviation after a plateau around the resonance frequency (inflection point at $f_1$).
\end{itemize}

\subsection{Experimental illustration}

Because of the important simplification of the model, a quantitative comparison between experimental and numerical results is not possible. However, a qualitative comparison is interesting, and proves that the dynamical system studied in this paper reproduces some key features of the behaviour of flute-like instruments.
\\

Experimentally, aeolian sounds occur for very low blowing pressures (which correspond to high values of the delay $\tau$), that is to say when the musician blows very gently in the instrument. It is why experimental observation of such sounds is particularly complicated.
The use of a pressure-controlled artificial mouth (which permits to play the instrument without musician, and to control very precisely the mouth pressure) provides new information about the dynamics of the instrument \cite{Ferrand_10}.
The simultaneous measurement of the mouth pressure $P_{alim}$ (related to the jet velocity $U_j$ by the Bernoulli equation for stationary flows) and the acoustic pressure under the labium  $P_{in}$ (see figure \ref{schema_coupe_flute}) allows to study the influence of some parameters on the characteristics of the oscillation regimes.
\\

\begin{figure}

\centering%
\includegraphics[width=\columnwidth]{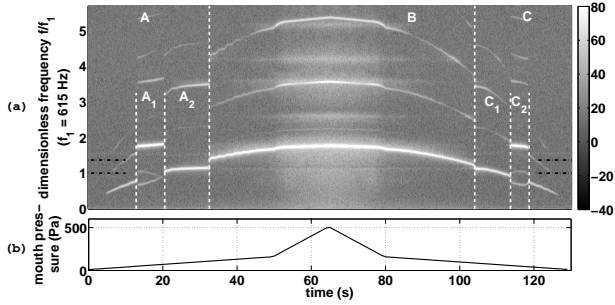}
\caption{Time-frequency representation (a) of the sound of a Zen-On Bressan plastic alto recorder played by an artificial mouth making an increasing and decreasing pressure ramp (b). The dark dot-dashed lines indicate the two first resonance frequencies ($f_1 \approx 615$ Hz and $f_2 \approx 842$ Hz) of the resonator. G sharp fingering (4th octave).}
\label{spectrogramme_se}
\end{figure}

\begin{figure}
\centering
\includegraphics[width=\columnwidth]{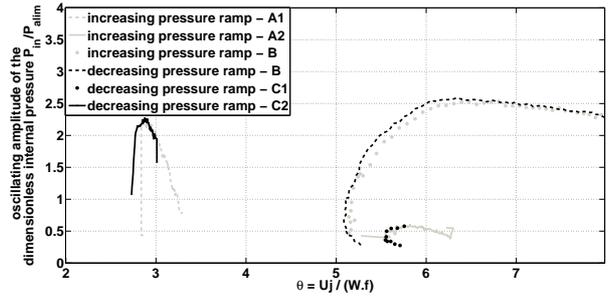}
\caption{Same experimental data as in figure \ref{spectrogramme_se}: representation of the dimensionless acoustic pressure amplitude measured in the resonator, with respect to the dimensionless jet velocity $\theta$. Letters in legend refer to letters in figure \ref{spectrogramme_se}-(a).} 
\label{Pbec_theta_se}
\end{figure}

Time-frequency analysis (figure \ref{spectrogramme_se}-(a)) of the sound produced by a Zen-On Bressan plastic alto recorder (previously studied by Lyons \cite{Lyons_81}) during both an increasing and a decreasing ramp of the mouth pressure (figure \ref{spectrogramme_se}-(b)) highlights several oscillation regimes (zones A and C) below the threshold of the principal regime corresponding to the expected note (zone B). Figure \ref{Pbec_theta_se} represents, for the same data, the dimensionless acoustic pressure amplitude $P_{in}/P_{alim}$ as a function of $\theta$. As for aeolian regimes on the numerical bifurcation diagram (figure \ref{figure_ampl_theta}), we observe that one of these regimes is located in a zone defined by $\theta < 4$, whereas the principal regime (zone B) appears for $\theta > 4$. If the second one (regime A2-C1 in figures \ref{spectrogramme_se}-(a) and \ref{Pbec_theta_se}) appears for higher values of $\theta$ (about $ 5.25 < \theta < 6.25$), it nevertheless can be considered as an aeolian regime. Indeed, as can be seen in figure \ref{spectrogramme_se}-(a), it corresponds to an oscillation on the first resonance mode of the pipe, whereas regimes A1-C2 and B correspond to oscillations on the second resonance mode. Consequently, the oscillation frequency of regime A2-C1 being lower than that of regimes A1-C2 and B, the value of $\theta$ is increased.

Moreover, for the two sounds which appear for low values of the mouth pressure, the bell-shaped evolution of the oscillating amplitude is clearly visible, recalling the characteristics of aeolian regimes of the model. On the contrary, the oscillating amplitude of the principal regime shows a different evolution, comparable to the one observed in figure \ref{figure_ampl_theta} for the standard regime.
As far as the frequency is concerned, large variations can be observed in figure \ref{spectrogramme_se}-(a), particularly close to the threshold of the standard regime B.
\\

The experimental observation of aeolian sounds on a recorder is especially interesting. Indeed, if these particular sounds are well-known for flue organ pipes (see for example \cite{Fletcher_76,Verge_97}), it is generally assumed that recorders can not operate on the 2nd (or higher) hydrodynamic mode of the jet, due to the small value of their ratio $W/h$ (with $W$ the mouth heigh and $h$ the channel exit heigh - see figure \ref{schema_coupe_flute}) \cite{Verge_97,Chaigne_Kergomard}. Indeed, this ratio is close to $4$ in recorder-like instruments whereas it can reach $12$ in flue organ pipes \cite{Verge_97}.
\\

Thereby, the very simple model studied here produces aeolian regimes, which present particular features which recall some experimental observations on a recorder played by an artificial mouth. Moreover, these observations on the amplitude and frequency evolution patterns for aeolian regimes can recall the results of Meissner in the case of a cavity resonator excited by an air jet \cite{Meissner_01}.

However, the confrontation between figures \ref{figure_ampl_theta} and \ref{Pbec_theta_se} highlights that the system under study produces aeolian regimes with a larger amplitude than those produced by a recorder. Such important differences between numerical and experimental results (in term of amplitudes) for low values of the dimensionless jet velocity $\theta$ have also been observed by Fletcher \cite{Fletcher_76} in the case of a flue organ pipe. It can probably be related to the very simple modeling of the complex behaviour of the air jet in flute-like instruments (see for example \cite{Fabre00,Coltman_92b,Verge_93,Fabre_96,Howe_81}), which is simply represented, in the system under study, by an amplification factor and a delay.


\section{Jumps between periodic solution branches}
\label{section_bif_br_perio}

In this section, we study system (\ref{eq:1st-order}) as in section \ref{section_sol_perio}, with the addition of a second acoustic mode (\textit{i.e.} two terms in equation (\ref{eq. admittance}), $m=2$). As for an open cylindrical resonator, the resonance frequency of the second mode is chosen close to twice the first resonance frequency \cite{Chaigne_Kergomard}.

\subsection{Register change and hysteresis phenomenon}

The use of a transfer function including two modes implies the existence of two standard regimes. Each of these regimes results from the coupling between one of the two acoustic modes and the first hydronamic mode of the jet ($n=0$).

\begin{figure}[h!]
\begin{minipage}[c]{.95\columnwidth}
\centering
\includegraphics[width=0.8\columnwidth]{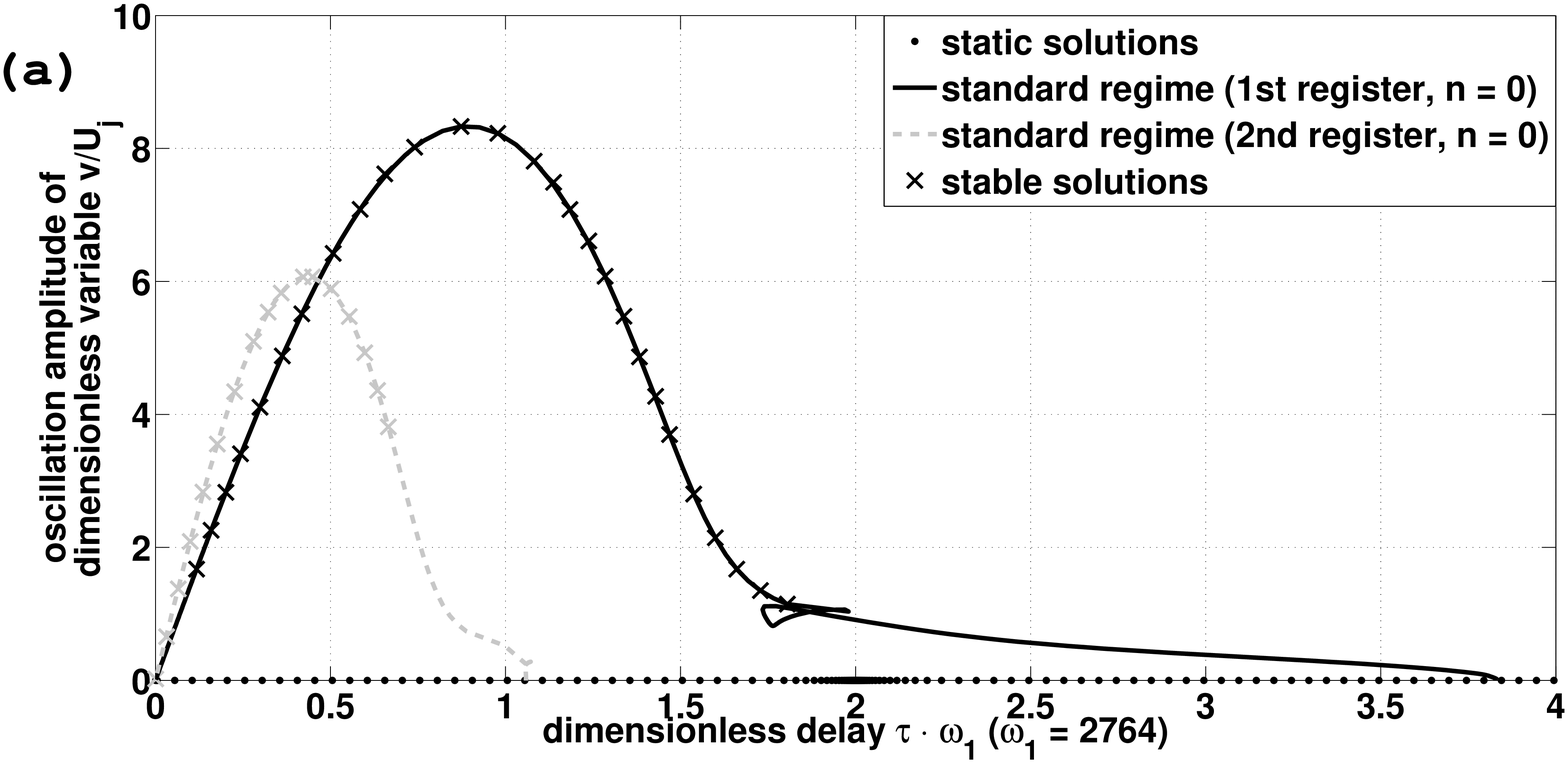}
\end{minipage}
\begin{minipage}[c]{.95\columnwidth}
\centering
\includegraphics[width=0.8\columnwidth]{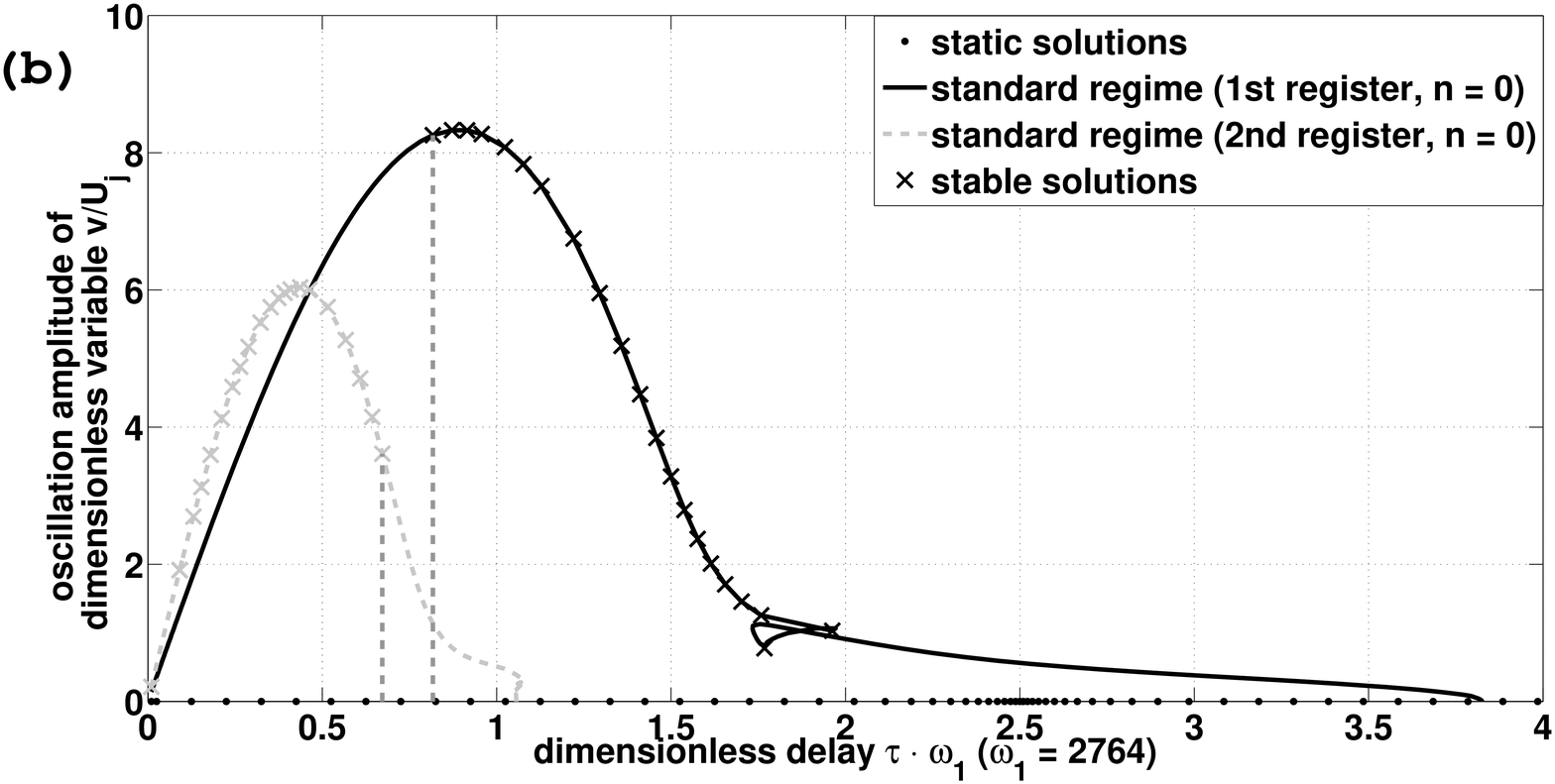}
\end{minipage}
\begin{minipage}[c]{.95\columnwidth}
\centering
\includegraphics[width=0.8\columnwidth]{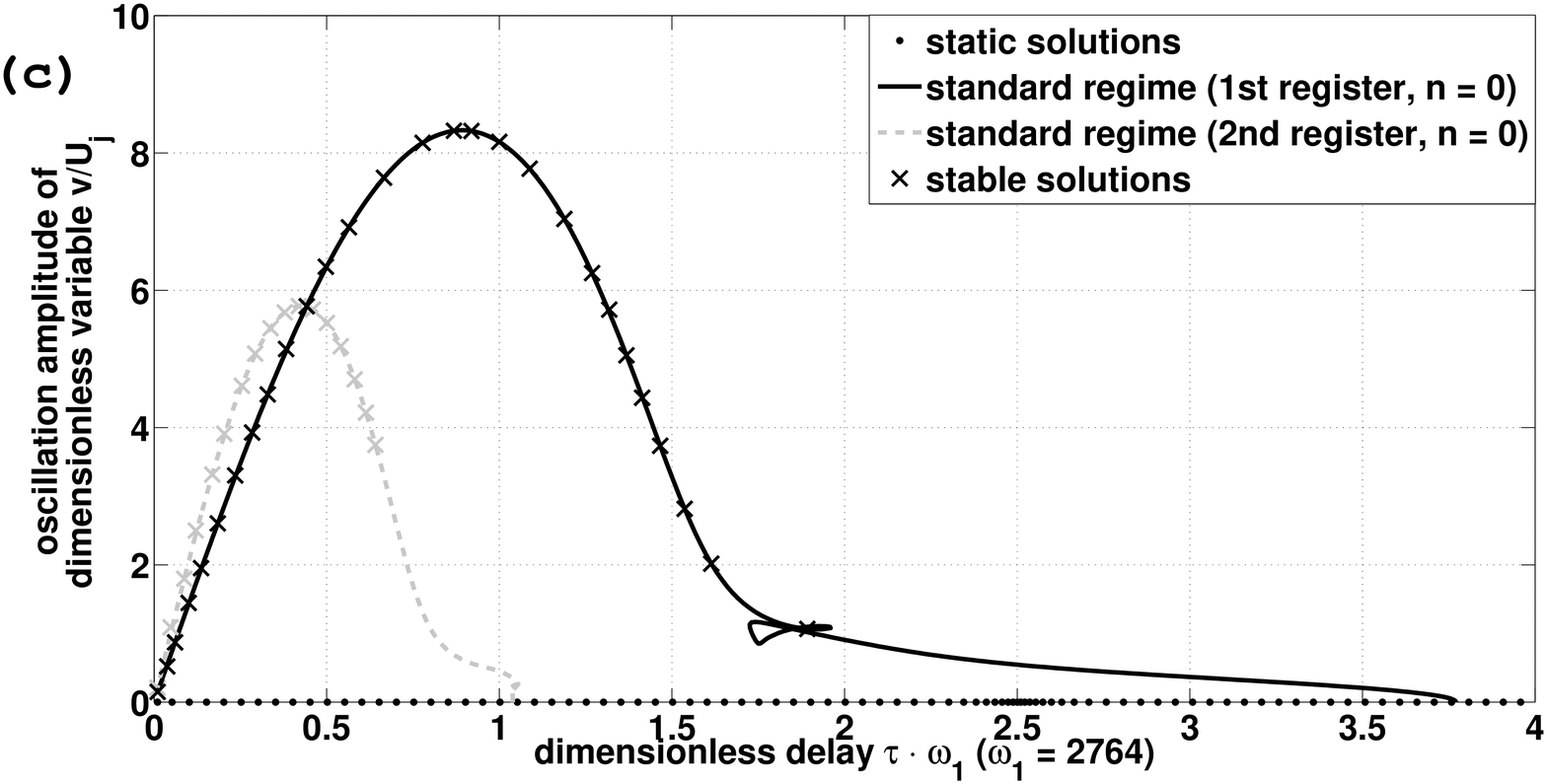}
\end{minipage}
\caption{Bifurcation diagrams of the system (\ref{eq:1st-order}), obtained with numerical continuation, for different values of inharmonicity of the two resonance modes. Parameter values: $m = 2$ ; $\alpha = 340$ ; $a_1 = 10$ ; $a_2 = 30$ ; $Q_1 = 100$ ; $Q_2 = 100$ ; $\omega_1 = 2764$.
(a) : $\omega_2 = 5510$, $\frac{\omega_2}{\omega_1} = 1.99$.
(b) : $\omega_2 = 5528$, $\frac{\omega_2}{\omega_1} = 2$. 
(c) : $\omega_2 = 5654$, $\frac{\omega_2}{\omega_1} = 2.05$. Abscissa: dimensionless delay $\tilde \tau = \omega_1 \tau$. Ordinate: oscillation amplitude of the dimensionless variable $v(\tilde t)/U_j$. Symbols $\times$ indicate the stable parts of the branches.}
\label{comparaison_diag_bif_inharm}
\end{figure}

For a second resonance frequency slightly lower than twice the first one, numerical continuation of the corresponding periodic solution branches leads to the bifurcation diagram shown in figure \ref{comparaison_diag_bif_inharm} - (a). As previously, it represents the oscillating amplitude of the dimensionless variable $v(\tilde t)/U_j$ as a function of the dimensionless delay $\tilde \tau$. Here again, the stability is addressed by examining the Floquet multipliers. However, there is a major difference with the case where only one mode of the resonator is considered: indeed, both stable and unstable portions appear (symbol $\times$ indicates the stable parts of the branches).

More precisely, we can distinguish three different zones in figure \ref{comparaison_diag_bif_inharm} - (a):
\begin{itemize}
\item{the first one is defined for $\tilde \tau > 0.7$; in this case the first standard regime (called "first register" in the context of musical acoustics) is stable, whereas the second standard regime ("second register") is unstable.}
\item{for $0.1 < \tilde \tau < 0.7$, the two standard regimes (\textit{i.e.} the two registers) are simultaneously stable.}
\item{in the third zone, where $\tilde \tau < 0.1$, the first register is unstable, and the second register is the only stable solution of the system.}
\end{itemize}

The existence of a range of the dimensionless delay $\tilde \tau$ (between $0.1$ and $0.7$) where two periodic solutions are simultaneously stable implies the existence of an hysteresis phenomenon. Indeed, starting from a value of $\tilde \tau$ where the first register is the only stable solution, and decreasing the delay $\tilde \tau$ (\textit{i.e.} "blowing harder"), we observe that the system follows the branch corresponding to the first register until it becomes unstable (for $\tilde \tau= 0.1$). At this point, the system synchronizes with the second register.

Starting from this new point, and increasing the delay $\tilde \tau$ (\textit{i.e.} "blowing softer"), the system follows the branch corresponding to the second register. It is only when this second branch becomes unstable (for $\tilde \tau = 0.7$) that the system comes back to the first register.

\begin{figure}[h!]
\centering
\includegraphics[width=\columnwidth]{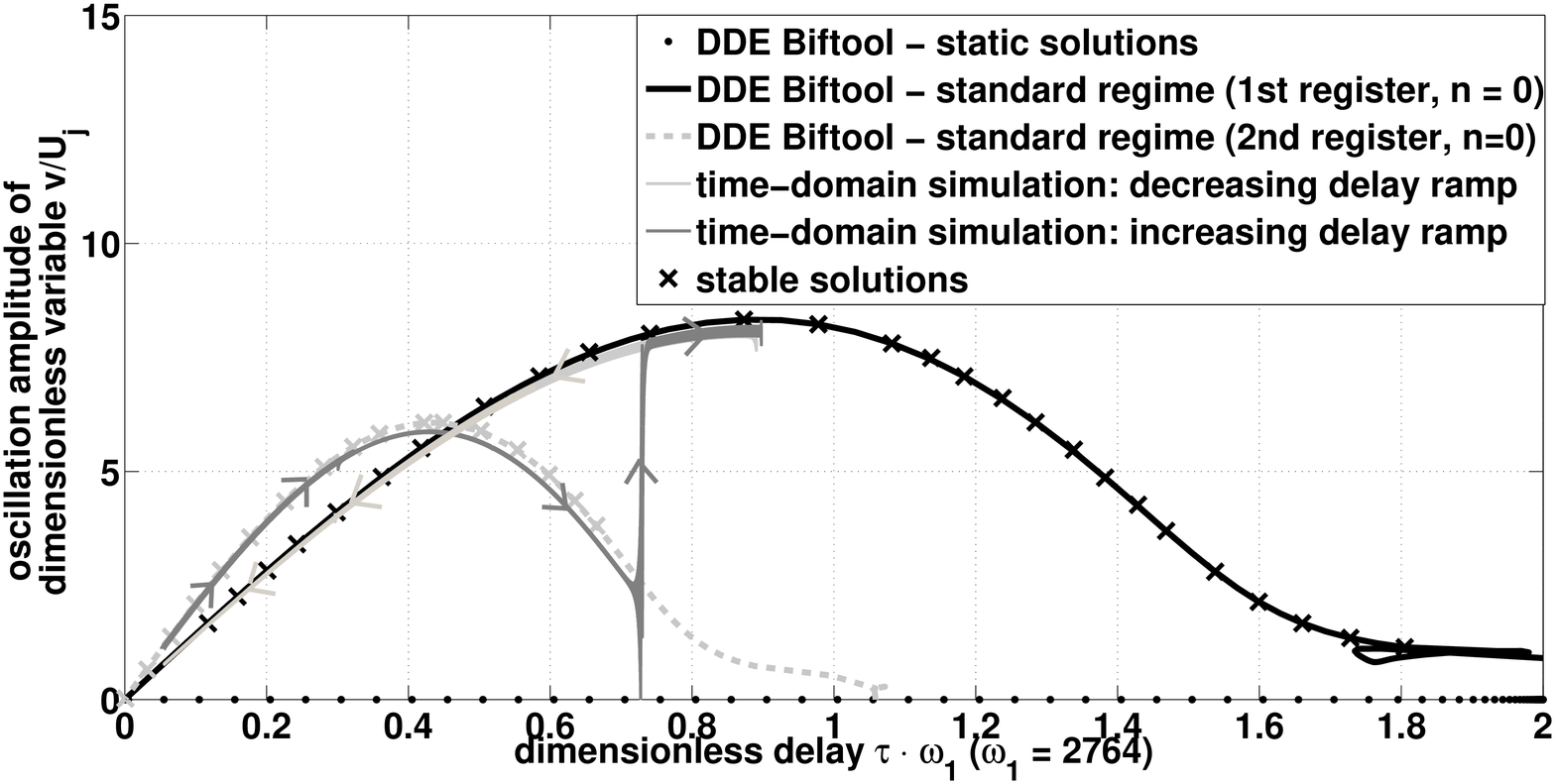}
\caption{Comparison between the bifurcation diagram of the system (\ref{eq:1st-order}), obtained with numerical continuation, and results provided by a time-domain solver. Same parameter values as in figure \ref{comparaison_diag_bif_inharm}-(a). Abscissa: dimensionless delay $\tilde \tau = \omega_1 \tau$. Ordinate: oscillating amplitude of the dimensionless variable $v(\tilde t)/U_j$. Symbols $\times$ indicate the stable parts of the branches.}
\label{comparaison_diag_bif_inharm_ZOOM}
\end{figure}

A time-domain simulation, using a Bogacki-Shampine method based on a third-order Runge-Kutta scheme \cite{ode23} allows to confirm and highlight this hysteresis phenomenon. In order to compare the different results, we superimpose, in figure \ref{comparaison_diag_bif_inharm_ZOOM} (which is a zoom of figure \ref{comparaison_diag_bif_inharm}-(a)), the bifurcation diagram and the oscillating amplitude computed with this time-domain solver, for both an increasing and a decreasing ramp of the delay $\tilde \tau$.

\subsection{Register change: experimental illustration}

\begin{figure}[h!]
\centering
\includegraphics[width=\columnwidth]{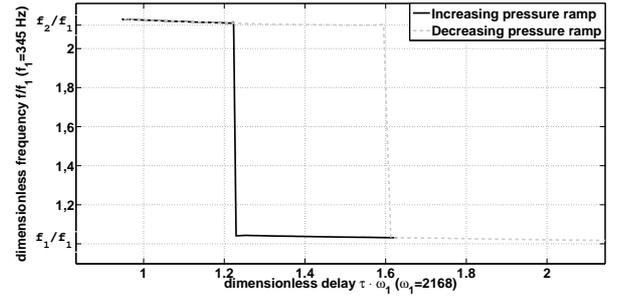}
\caption{Dimensionless playing frequency $f/f_1$ of a Zen-On Bressan plastic alto recorder blown by an artificial mouth, during an increasing (solid line) and decreasing (dashed line) blowing pressure ramp, showing jumps between the two standard regimes. Two first resonance frequencies of the resonator: $f_1 = 345$ Hz, $f_2 = 733$ Hz. F fingering (third octave).} 
\label{hysteresis_BA}
\end{figure}

In flute-like instruments, it is well-known by musicians that blowing harder in the instrument eventually leads to a "jump" to the note an octave above.
Figure \ref{hysteresis_BA} represents the frequency evolution of the sound produced by a Zen-On Bressan plastic alto recorder played by the artificial mouth, during an increasing and a decreasing ramp of the alimentation pressure.

These experimental results illustrate the register change phenomenon, corresponding to the "jump" from a periodic solution branch of frequency $f_1$ to another of frequency $f_2 \approx 2 \cdot f_1$.
\\

An hysteresis is also observed. Recalling that in flute-like instruments, the blowing pressure is related to the convection delay $\tau$ of the hydrodynamic perturbations along the jet, these experimental results can be compared to numerical results presented in figure \ref{comparaison_diag_bif_inharm_ZOOM}. It shows again that the system under study reproduces classical behaviour of recorder-like instruments.




\section{Influence of the ratio of the two resonance frequencies}
\label{section_reg_QP}

Numerical continuation allows a systematic investigation of the parameters influence on the oscillation regimes.
In this section, we focus on the influence of the ratio of the two resonance frequencies. Indeed, this parameter is well-known to instrument makers and players for its influence on the sound characteristics \cite{Bolton}. Moreover, its influence has been proved for other musical instruments (for example, in \cite{Dalmont_95}).

\subsection{Stability of periodic solution branches}

Figures \ref{comparaison_diag_bif_inharm} present bifurcation diagrams obtained for different values of this ratio (respectively $\frac{\omega_2}{\omega_1} = 1.99$ ; $\frac{\omega_2}{\omega_1} = 2$ and $\frac{\omega_2}{\omega_1} = 2.05$). Only the second resonance frequency $\omega_2$ varies from one case to the other; all the others parameters are kept constant.

For all these bifurcation diagrams, we found an aeolian regime (see section \ref{section_reg_periodiques}) in a zone defined by about $0.8 < \tilde \tau <5.5$ . For sake of readability, the corresponding curves are not represented here. The dark gray curve corresponds to the first register, the light gray dashed one corresponds to the second register, and symbols $\times$ indicate stable parts of the branches.

Although the resonance frequency of the second mode is only slightly modified from one case to the other, one observes significant changes in the stability properties of the two registers.
In the case of two perfectly harmonic acoustic modes ($\omega_2 = 2 \omega_1$), represented in figure \ref{comparaison_diag_bif_inharm} - (b), the first register is initially stable, and becomes unstable when the delay $\tilde \tau$ decreases (\textit{i.e.} when the blowing pressure increases). On the contrary, the second register is first unstable and becomes stable when the delay $\tilde \tau$ decreases. For the range of $\tilde \tau$ located between the loss of stability of the first register and the stabilization of the second register (\textit{i.e.} between the two vertical dot-dashed lines in figure \ref{comparaison_diag_bif_inharm} - (b)), there is no stable solution, neither static, nor periodic.

\begin{figure}[h!]
\centering
\includegraphics[width=\columnwidth]{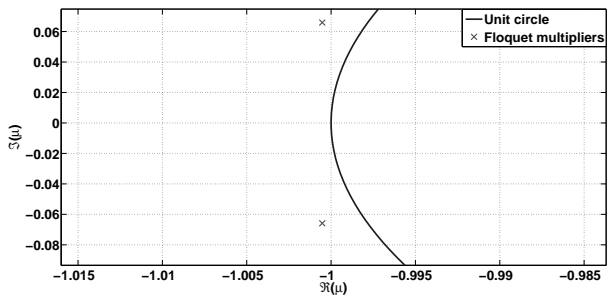}
\caption{Representation in the complex plane of the Floquet multipliers for a point located just before the stabilization of the second periodic solution branch (related to the second register), at $\tilde \tau = 0.68$. Parameter values are the same as in figure \ref{comparaison_diag_bif_inharm} - (b).}
\label{figure_mult_Floquet}
\end{figure}

In order to determine the nature of the different resulting bifurcations, we propose to study Floquet multipliers in the vicinity of stability changes. As shown in figure \ref{figure_mult_Floquet}, stabilization of the second register occurs, at $\tilde \tau = 0.68$, by the crossing of the unit circle by two conjugate complex Floquet multipliers.

\begin{figure}[h!]
\centering
\includegraphics[width=\columnwidth]{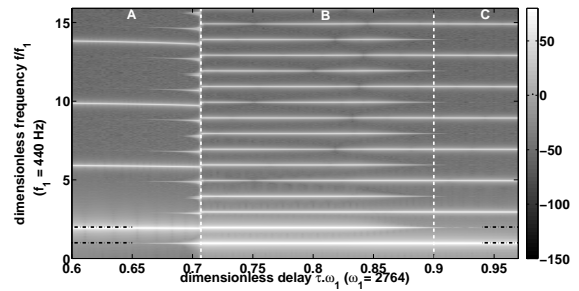}
\caption{Time-frequency representation obtained by a time-domain simulation of the system (\ref{eq:1st-order}), using a third-order Runge-Kutta scheme. A slowly increasing ramp of the dimensionless delay $\tilde \tau = \omega_1 \tau$ is achieved. Zone $A$ corresponds to the second register, zone $C$ corresponds to the first register, and zone $B$ to a quasiperiodic regime. Parameter values are the same as in figure \ref{comparaison_diag_bif_inharm} - (b). The dark dot-dashed lines indicate the two first resonance frequencies of the transfer function ($f_1 = 440$ Hz and $f_2 = 880$ Hz).}
\label{spectro_simulink_QP}
\end{figure}

\noindent
In the case of a direct bifurcation, this Hopf bifurcation of the limit cycle leads to a quasiperiodic regime \cite{Nayfeh}. The numerical tools used here do not allow numerical continuation of this kind of solutions. On the other hand, time-domain integration methods allow to study such regimes. Figure \ref{spectro_simulink_QP} shows the time-frequency analysis of the signal obtained with the same time-domain solver as previously, by increasing the dimensionless delay $\tilde \tau$ in a quasi-static way from an initial value for which the second register is stable, to a final value for which the first register is stable. We note three different regimes: zones A ($\tilde \tau < 0.71$) and C ($\tilde \tau > 0.9$) correspond respectively to the second and first registers. In zone B ($0.71 < \tilde \tau < 0.9$), the presence of multiple frequencies reveals the existence of a quasiperiodic regime, which agrees with the results of Floquet analysis (presented in figures \ref{comparaison_diag_bif_inharm} - (b) and \ref{figure_mult_Floquet}).
\\

As shown in figure \ref{comparaison_diag_bif_inharm} - (a), a small decrease of the second mode resonance frequency (of about 0.6 \%) compared to the harmonic case $\omega_2 = 2\omega_1$ does not alter the general form of the bifurcation diagram: the first register is stable for high values of $\tilde \tau$, whereas the second register is first unstable and becomes stable when the delay $\tilde \tau$ decreases. However, the stability ranges are significantly modified. Compared to the previous case, the first register is stable for smaller values of the delay $\tilde \tau$, which leads to the existence of a range of $\tilde \tau$ where the two registers are simultaneously stable. In the later case, the hysteresis phenomenon emphasized in section \ref{section_bif_br_perio} becomes possible.
\\

These stability ranges are affected again by an increase of the second mode resonance frequency (leading to a difference of about 2\% between $\omega_2$ and $2 \omega_1$). Figure \ref{comparaison_diag_bif_inharm} - (c) shows the resulting bifurcation diagram. One can notice that the first register is now stable for all values of $\tilde \tau$ between $2$ and $0$.
Hence, once the system is synchronized on this branch of solutions,  no register change is possible, and the only way to make the system oscillate on the second register is to change initial conditions. Time-domain simulations results (not presented here) show good agreement with these characteristics.
\\


\subsection{Experimental illustration}

As emphasized in section \ref{Intro}, the simplicity of the studied system prevents a quantitative comparison between numerical results and experimental datas. However, we can qualitatively compare some experimental phenomena with numerical results presented in the previous subsection.
\\

The use of an artificial mouth with a real instrument highlights the influence of the inharmonicity on the sound characteristics. Indeed, the measured inharmonicity depends on the fingering.

\begin{figure}[h!]
\centering%
\includegraphics[width=\columnwidth]{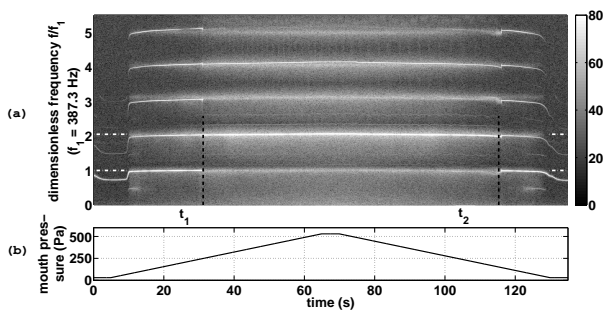}
\caption{Time-frequency representation (a) of the sound of a Yamaha (YRA-28BIII) plastic alto recorder played by an artificial mouth during an increasing and decreasing blowing pressure ramp (b). The white dot-dashed lines indicate the two first resonance frequencies ($f_1 \approx 387.3 $ Hz, $f_2 \approx 793.3 $ Hz) of the resonator. The dark dashed lines indicate register changes. G fingering (3th octave), with a slightly positive inharmonicity ($\frac{\omega_2}{\omega_1} \approx 2.04 $).}
\label{spectro_harm}
\end{figure}

Thus, for a small positive inharmonicity ($\frac{\omega_2}{\omega_1} \approx 2.04$), an increasing mouth pressure ramp (corresponding to a decrease of the convection delay $\tau$) causes a regime change from the first register to the second register, including an hysteresis effect (as shown in figure \ref{spectro_harm}).

\begin{figure}[h!]
\centering
\includegraphics[width=\linewidth]{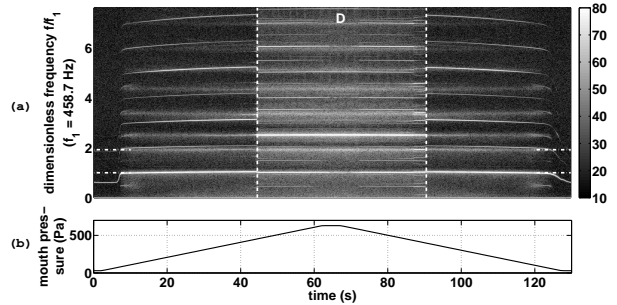}
\caption{Time-frequency representation (a) of the sound of a Yamaha (YRA-28BIII) plastic alto recorder played by an artificial mouth during an increasing and decreasing blowing pressure ramp (b). The white dot-dashed lines indicate the two first resonance frequencies ($f_1 \approx 458.7 $ Hz and $f_2 \approx 883.3 $ Hz) of the resonator. B-flat fingering (3th octave), with a significant negative inharmonicity ($\frac{\omega_2}{\omega_1} \approx 1.92 $).}
\label{spectro_inharm}
\end{figure}

\noindent
For a larger negative inharmonicity of the resonance frequencies ($\frac{\omega_2}{\omega_1} \approx 1.92 $), the same experiment leads to a very different behaviour. As shown in figure \ref{spectro_inharm}, we observe a transition from the first register to a quasiperiodic regime (in zone D) without hysteresis effect. 
\\

Comparison of these results with those of figures \ref{comparaison_diag_bif_inharm} - (a), \ref{comparaison_diag_bif_inharm} - (b) and \ref{comparaison_diag_bif_inharm} - (c) proves that the behaviour of both the studied system and the real instrument strongly depends on the inharmonicity of the two first resonance frequencies.

A small change of this parameter value is enough to alter the oscillation thresholds of the different regimes, their stability properties, or even the nature of the oscillation regimes.
Depending on the case, the system can "jump" between branches of periodic solutions (with hysteresis effect), or jump from a periodic solution branch to a quasiperiodic regime. 

Such quasiperiodic sounds were previously observed experimentally on flute-like instruments and obtained through numerical simulations, for example by Fletcher \cite{Fletcher_76} and Coltman \cite{Coltman_06}. Basing on the comparison between passive resonance frequencies of the pipe and sounding frequencies of the instrument, Coltman \cite{Coltman_06} explained this phenomenon by a beat between two frequencies: a first one corresponding to an harmonic (exact multiple of the fundamental frequency), and a second corresponding to an oscillation on an higher resonance mode of the pipe. Our numerical and experimental observations, highlighting that a slight shift of the second resonance frequency is enough to make the system (respectively the instrument) produce quasiperiodic regimes, show good agreement with such observations.
\\

Moreover, these results seem to be consistent with the experience of a recorder maker \cite{Bolton}: according to him, the first register is stable for a wider range of alimentation pressure in the case of strong inharmonicity and the case of a perfect harmonicity should be avoided to prevent instabilities. It is interesting to note that this feature is an important difference between recorders and transverse flutes. Indeed, unlike recorder makers, flute makers seek a good harmonicity to allow flutists to play in tune on different octaves using a same fingering.


\section{Conclusion}

We have studied a nonlinear delay dynamical system with small number of degrees of freedom. We focused on periodic solutions. Indeed, this system is inspired by flute-like instruments and in most cases, notes produced by these instruments are periodic oscillations.
\\

Because of the drastic simplifications, the studied system can not be considered \textit{a priori} as a model for these instruments. However, as shown in this paper, the use of numerical continuation tools, providing a more global vision of the dynamics of the system, highlights that the system not only presents a considerable variety of periodic regimes, but also has a second interest. Indeed, it qualitatively reproduces phenomena observed experimentally on flute-like instruments such as amplitude and frequency evolutions for both standard and aeolian regimes, regime changes with hysteresis effect, and quasiperiodic oscillations.
\\

We can furthermore investigate the influence of some parameters related to instrument maker's issues. We focused here on the role of the inharmonicity of the two first resonance frequencies. Analysis of bifurcation diagrams leads to results that are consistent with the empirical knowledge of an instrument maker.
\\

However, it would be hazardous to use the studied dynamical system as a predictive tool, for example for the design of musical instruments. Indeed, some parameters of the system can not be related to mesurable physical quantities. Moreover, as we emphasized in section \ref{Intro}, some important physical phenomena have not been taken into account. These two elements make vain any attempt of quantitative comparison between numerical and experimental results.
Particularly, we did not discussed about the sound timbre, which corresponds to the spectrum of the sound produced by the instrument. This characteristics is essential in a musical context since it allows us to distinguish, in the case of a steady-state regime, the sound of a flute from that, for example, of a trumpet. 
The system studied here can not be used to predict the sound timbre, since some elements known for their significant influence on the spectrum (but also on the amplitude of the different regimes) have been neglected:
\begin{itemize}
\item{the offset between the position of the labium and the channel exit. Due to symmetry properties of the nonlinear function, this parameter controls the ratio between even and odd harmonics \cite{Fletcher_80,Coltman_06,Verge_97}.}
\item{nonlinear losses due to flow separation and vortex shedding at the labium, which have an important influence on the sound level \cite{Howe_75,Hirschberg_95,Fabre_96,Verge_97,Auvray_12}.}
\item{the dipolar character of the pressure source, created by the oscillation of the jet around the labium \cite{Rayleigh,Fabre00,Coltman_76}.}
\item{the jet velocity profile \cite{Fabre00,Verge_94}, and an accurate modeling of the behaviour of the unstable jet (\cite{Nolle_98,Coltman_92b} for example), particularly for low values of the dimensionless jet velocity $\theta$ \cite{Hirschberg_95, Fabre00}.}
\end{itemize}

Taking into account these two first elements do not compromise the use of the approach presented in this paper. Only the number of parameters and degree of freedom (and thus the computation time) increases. However, taking into account the third element involves the presence of a delayed derivative term in the right-hand side of the second equation of system (\ref{systeme_eq_modele}). This change transforms the delay system into a neutral delay dynamical system, and thus introduces additional difficulties in numerical continuation and system analysis \cite{Barton_06}.

\section{Acknowledgements}
The authors would like to thank Philippe Bolton for interesting and helpful discussions during the completion of this work.


\newpage

\appendix

\section{DDE Biftool: Reformulation of the studied system}
\label{annexe_Biftool}

To be implemented in the software DDE Biftool, system (\ref{systeme_eq_modele}) needs to be reformulated as follows:

\begin{equation}
\mathbf{\dot x}(t) = f(\mathbf{x}(t),\mathbf{x}(t-\tau),\boldsymbol{\gamma})
\label{eq_exple_syst_dyn}
\end{equation}

\noindent 
where $\tau$ is a delay, $\mathbf{x}$ the vector of state variables, and $\boldsymbol{\gamma}$ the set of parameters.
\\

\noindent
We detail here such a reformulation in the case of a single-mode transfer function (equation (\ref{eq. admittance})):
\begin{equation}
Y(\omega) = \frac{\mathrm{j}\omega \cdot a_1}{\omega_1 ^2 - \omega ^2 + \mathrm{j}\omega \frac{\omega_1}{Q_1}}
\label{admittance 1 mode}
\end{equation}
where $\omega$ is the pulsation, and $a_1$, $\omega_1$ and $Q_1$ are respectively the modal amplitude, the resonance pulsation and the quality factor of the resonance mode. 
\\

\noindent
Let us inject equation (\ref{admittance 1 mode}) in the third equation of system (\ref{systeme_eq_modele}):

\begin{equation}
V(\omega) \cdot \left[\omega_1 ^2 - \omega ^2 + \mathrm{j}\omega \frac{\omega_1}{Q_1}\right] = \mathrm{j} \omega \cdot a_1 \cdot P(\omega).
\end{equation}

\noindent
An inverse Fourier transform leads to:
\begin{equation}
\ddot v(t) + \omega_1^2 \cdot v(t) + \frac{\omega_1}{Q_1} \cdot \dot v(t) = a_1 \cdot \dot p(t).
\label{eq_temp}
\end{equation}

\noindent
The right-hand side of this expression can be developed using the second equation of system (\ref{systeme_eq_modele}):
\begin{equation}
p (t) = \alpha \cdot \tanh \left[z(t)\right].
\end{equation}

\noindent
Differentiating with respect to time, and using the first equation of system (\ref{systeme_eq_modele}), we obtain:
\begin{equation}
\begin{split}
\dot p (t) &= \alpha \cdot \dot z(t) \cdot \{1-\tanh^2[z(t)]\} \\
& = \alpha \cdot \dot v(t-\tau) \cdot \{1-\tanh^2[v(t-\tau)]\}
\end{split}
\end{equation}

\noindent
Injecting this expression in equation (\ref{eq_temp}) leads to:
\begin{equation}
\begin{split}
\ddot v(t) + \frac{\omega_1}{Q_1}& \dot v(t) + \omega_1 ^2 v(t) = a_1 \alpha \cdot \dot v(t-\tau)  \\
& \{ 1-\tanh^2 [v(t-\tau)] \}
\end{split}
\label{eq_temporelle}
\end{equation}

\noindent
To improve numerical conditioning of the problem, it is convenient to make the temporal variables dimensionless. Let us introduce the new variables:
\begin{System*}
\tilde{t} = \omega_1 \cdot t\\
\tilde{\tau} = \omega_1 \cdot \tau%
\end{System*}%

\noindent
Equation (\ref{eq_temporelle}) becomes:

\begin{equation}
\begin{split}
& \frac{\deriv^2 v(\tilde{t})}{\deriv(\frac{\tilde{t}}{\omega_1})^2} +  \frac{\omega_1}{Q_1} \cdot \frac{\deriv(v(\tilde{t}))}{\deriv(\frac{\tilde{t}}{\omega_1})} + \omega_1^2 \cdot v (\tilde{t}) \\
& = 
a_1 \alpha \frac{\deriv[v(\tilde{t}-\tilde{\tau})]}{\deriv(\frac{\tilde{t}}{\omega_1})} 
\cdot \{ 1-\tanh^2[v(\tilde{t}-\tilde{\tau})]   \}.
\end{split}
\end{equation}

\noindent
It leads to:

\begin{equation}
\begin{split}
\ddot v(\tilde{t}) + \frac{1}{Q_1} & \cdot \dot v(\tilde{t}) + v(\tilde{t}) = \frac{a_1 \alpha}{\omega_1} \cdot \dot v(\tilde{t}-\tilde{\tau}) \\
&\{1-\tanh^2\[v(\tilde{t}-{\tilde{\tau}})\]\} ,
\end{split}
\end{equation}

\noindent
where $\dot v$ now define the derivative of $v$ with respect to dimensionless time $\tilde{t}$.

\noindent
We define a new variable:
\begin{equation}
y (\tilde{t}) = \dot v (\tilde{t}),
\end{equation}

\noindent
which leads to the correct form of the system (given by equation (\ref{eq_exple_syst_dyn})):
\begin{System}
\begin{split}
\dot v(t)= & y(t)\\
\dot y (\tilde{t}) = &  \frac{a_1 \alpha}{\omega_1} \cdot y(\tilde{t}-\tilde{\tau})  \cdot \{1-\tanh^2\[v(\tilde{t}-\tilde{\tau})\]\} \\
& - v(\tilde{t}) - \frac{1}{Q_1} y(\tilde{t}).
\end{split}
\end{System}

\vspace{1cm}
Generalization to a system including $m$ resonance modes is straightforward. In this case, equation (\ref{eq. admittance}) contains $m$ terms (with $m > 1$):

\begin{equation}
Y = \sum_{k=1}^m \frac{\mathrm{j}\omega \cdot a_k}{\omega_k ^2 - \omega ^2 + \mathrm{j}\omega \frac{\omega_k}{Q_k}}
\end{equation}
\noindent
and it leads to a system of $2m$ equations (\textit{i.e.} 2 equations by resonance mode), of the following form:

\begin{System}
\begin{split}
\forall i=&[1, 2 ..., m]: \\
\dot v_{i}(\tilde{t}) =& y_i(\tilde{t})\\
\dot y_i (\tilde{t}) =& \frac{a_i \alpha}{\omega_1} \{ 1-\tanh^2\left[ \sum_{k=1}^m v_k(\tilde{t}-\tilde{\tau}) \right] \} \\
& \sum_{k=1}^m \left[ y_k(\tilde{t}-\tilde{\tau}) \right] - \left( \frac{\omega_i}{\omega_1} \right)^2 v_{i}(\tilde{t}) \\
&- \frac{\omega_i}{\omega_1 Q_i} y_i(\tilde{t}).  
\end{split}
\end{System}

\section{Linear analysis around the equilibrium solution}
\label{annexe_AL}

\subsection{Eigenvalues of the Jacobian}
Linearisation of system (\ref{eq:1st-order}) around equilibrium solution (\ref{equilibrium_sol}) allows to compute the eigenvalues $\lambda$ of the Jacobian matrix of the studied system along the branch of equilibrium solutions. Their real parts Re($\lambda$) determine the stability properties of the static solution: it is stable if and only if all the eigenvalues have negative real parts \cite{Nayfeh}.

\begin{figure}[h!]
\centering
\includegraphics[width=\columnwidth]{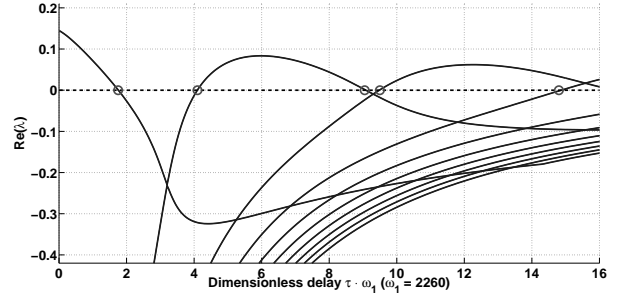}
\caption{Real part of the eigenvalues $\lambda$ of the Jacobian matrix of system (\ref{eq:1st-order}) linearised around static solution (\ref{equilibrium_sol}), with respect to the dimensionless delay $\tilde \tau = \tau \cdot \omega_1$. Hopf bifurcations (intersections with the horizontal line  Re($\lambda$)= $0$) are highlighted by circles. Parameters value: $m = 1$, $\alpha = 10$ ; $a_1 = 70$ ; $\omega_1 = 2260$ ; $Q_1 = 50$.}
\label{figure_Re_jacobien}
\end{figure}

Considering, for sake of clarity, a single acoustic mode of the resonator (\textit{i.e.} $m = 1$ in equation (\ref{eq. admittance})), we represent, in figure \ref{figure_Re_jacobien}, Re($\lambda$) with respect to the continuation parameter $\tilde \tau$, along the branch of equilibrium solutions. 
This representation highlights that this static solution is stable for $1.8 < \tilde \tau < 4.1$ and for $9.1 < \tilde \tau  < 9.5$. Analysis of both real and imaginary parts of the eigenvalues shows that the equilibrium solution is a focus point when it is stable, and a saddle point when it is unstable \cite{Nayfeh}.
Moreover, inspection of the intersection points with the x-axis (defined by Re($\lambda$) = 0) determines Hopf bifurcations \cite{Nayfeh}, which correspond to the birth of the different periodic solution branches (here highlighted by circles at $\tilde \tau  = 1.8$ ; $\tilde \tau  = 4.1$ ; $ \tilde \tau  = 9.1$ and $\tilde \tau  = 9.5$). 

\subsection{Open-loop gain}

In a feedback loop system such as the one presented in section \ref{section_systeme}, another approach to get information about the destabilization mechanisms of the equilibrium solution is to study the open-loop gain $G_{ol} (\omega)$ of the linearised system. Indeed, the emergence of auto-oscillations from the equilibrium solution (corresponding to a Hopf bifurcation) is possible under the two following conditions \cite{Fletcher_93}:

\begin{itemize}
\item{modulus $G$ of the linearised open-loop gain $G_{ol}$ must be equal to 1.}
\item{phase $P$ of the linearised open-loop gain $G_{ol}$ must be a multiple of $2 \pi$.}
\end{itemize}

Writting system (\ref{systeme_eq_modele}) in the frequency domain leads to: 
\begin{System}
Z (\omega) = V(\omega) \exp^{-\mathrm{j}\omega \tau}\\
P (\omega) = \alpha \cdot \tanh(Z(\omega))\\
V (\omega) = Y(\omega) \cdot P(\omega)
\end{System}

\noindent
and linearisation of the second equation around the equilibrium solution (\ref{equilibrium_sol}) leads to the open-loop gain:

\begin{equation}
G_{ol} (\omega) = \alpha Y(\omega) \cdot e^{-\mathrm{j} \omega \tau},
\end{equation}

\noindent
The conditions of emergence of auto-oscillations are then given by:
\begin{System}
G(\omega) = \alpha \cdot |Y(\omega)| = 1\\
P(\omega) = arg(Y(\omega)) - \omega \tau = -n \cdot 2\pi
\label{equation_GBO}
\end{System}
 
\noindent
where $n$ is an integer.
\\

\begin{figure}[h!]
\centering
\includegraphics[width=\columnwidth]{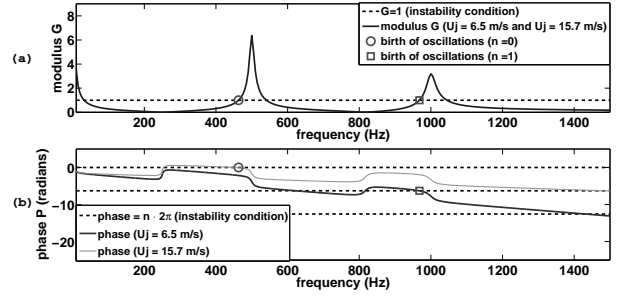}
\caption{Open-loop gain modulus G (a) and phase P (b) defined by eq. (\ref{equation_GBO}), for two different values of the jet velocity $U_j$. Emergence of self-sustained oscillations is possible if the modulus is equal to one (dash line, in (a)), and if the phase crosses a straight line with equation $P= -n \cdot 2 \pi$ (dash lines, in (b)).}
\label{diagramme_bode}
\end{figure}

To exemplify the conditions given by equation (\ref{equation_GBO}), let us consider a transfer function $Y(\omega)$ representing the first two modes of a cylindrical resonator ($m=2$ in equation (\ref{eq. admittance})). Figure \ref{diagramme_bode} shows the variables $G$ (a) and $P$ (b) with respect to $\omega$. Two different values of the delay $\tau$ are considered (corresponding to  two different values of the jet velocity $U_j=6.5ms^{-1}$ and $U_j=15.7ms^{-1}$) for phase $P$ ($G$ is independent of $\tau$). The points where equations (\ref{equation_GBO}) are fulfilled are marked with circles ($\circ$) when $n=0$ and squares ($\square$) when $n=1$. Plots of $P$ for lower values of $U_j$ (\textit{i.e.} for higher values of $\tau$) would have revealed other intersections with horizontal lines $P=-n \cdot 2\pi$, with larger values of $n$.

Provided the amplification $\alpha$ is large enough, this example highlights the existence of an infinity of solutions of equations (\ref{equation_GBO}) for different values of $U_j$, each solution being related to a given value of the integer $n$. Therefore, for each value of $n$, an instability may emerge from the equilibrium branch.	

From the physics point of view, the integer $n$ represents the rank of the hydrodynamic mode of the jet involved in the emergence of the instability. 



\newpage
\bibliographystyle{ieeetr}
\bibliography{biblio_toymodel}

\end{document}